\documentstyle[preprint,prd,aps,floats,eqsecnum,epsf]{revtex}
\newcommand{\nwl}{\\[1mm]}
\newcommand{\edc}{\end{document}}
\newcommand{\bb} {\begin{thebibliography}}
\newcommand{\eb}{\end{thebibliography}}
\newcommand{\bi}[1]{\bibitem{#1}}
\newcommand{\bc}{\begin{center}}
\newcommand{\ec}{\end{center}}

\newcommand{\be}{\begin{equation}\small}
\newcommand{\ee}{\end{equation}\normalsize}
\newcommand{\bea}{\begin{eqnarray}}
\newcommand{\eea}{\end{eqnarray}}
\newcommand{\ba}{\begin{array}{l}   }
\newcommand{\lab}[1]{\label{#1}}
\newcommand{\ea}{\end{array}}

\newcommand{\dsfrac}{\displaystyle\frac}
\newcommand{\ds} {\displaystyle}

\newcommand{\re}[1]{(\ref{#1})}

\newcommand{\ci}{\cite}

\newcommand{\dsint}{\ds\int}

\newcommand{\hash}{\hbar }
\newcommand{\hc}{\hash c }

\newcommand{\intdf} {\ds{\int  \cal{D} \phi } }
\newcommand{\intbes} {\dsint _{0}^{\infty} }

\newcommand{\veff}{V_{\mbox{eff}}}

\newcommand{\cl}{  \ell     }
\newcommand{\dd}{\partial}

\newcommand{\half}{\frac{1}{2}}
\newcommand{\mnol}{m_{0}^{2}}

\newcommand{\delnol}{\Delta_{0}^{2}}
\newcommand{\om}{\Omega^2}

\newcommand{\omb}{{\bar{\Omega}} }
\newcommand{\delb}{{\bar{\Delta}} }

\newcommand{\inolm}{I_0(\Omega)}
\newcommand{\inold}{I_0(\Delta)}

\newcommand{\phiz}{\phi_0}

\newcommand{\tochka}{  .}
\newcommand{{\vergul}}{  ,}
\newcommand{\vecA}{\vec{A}}
\newcommand{\phizb}{\bar{\phi}_0}
\newcommand{\mc}{m_{c}^{2} }
\newcommand{\veps}{\varepsilon }
\newcommand{\lna}{\ln\alpha}
\newcommand{\lnaa}{\ln^2\alpha}
\newcommand{\lnx}{\ln x_1}
\newcommand{\lnom}{\ln\dsfrac{\mu^2}{\omb^2}}
\newcommand{\lndel}{\ln\dsfrac{\mu^2}{\delb^2}}

\newcommand{\onehalfspace}
{\renewcommand{\baselinestretch}{1.2}\large\normalsize}

\onehalfspace
\begin{document}
\sloppy
\draft
\title{   \Large\bf {  Ginzburg Landau
theory of superconductivity at fractal dimensions}}
\author
{ Chul  Koo Kim\thanks{E-mail: ckkim@phya.yonsei.ac.kr}, A. Rakhimov \thanks
{ E-mail:  rakhimov@rakhimov.ccc.uz.  Permanant address: Institute of Nuclear Physics, Tashkent,
    Uzbekistan (CIS)}  and
 Jae Hyung Yee \thanks{E-mail: jhyee@phya.yonsei.ac.kr}
}

\address{Institute of Physics and Applied Physics, Yonsei
University, Seoul, 120-749, Korea\\}
\maketitle
\begin{abstract}

 The post  Gaussian effective potential
 in $D=2+2\veps$ dimensions is evaluated
 for the Ginzburg-Landau
 theory  of superconductivity.
Two and three loop integrals for the post Gaussian correction terms
in $D=2+2\veps$ dimensions are calculated and $\veps$-expansion for
these integrals are constructed. In $D=2+2\veps$ fractal dimensions
Ginzburg Landau parameter turned out to be sensitive to $\veps$ and
the contribution of the post Gaussian term is larger than that for
$D=3$. Adjusting $\veps$ to the recent experimental data on $\kappa
(T)$ for high -$T_c$ cuprate superconductor $Tl_2Ca_2Ba_2Cu_3O_{10}
(T\cl-2223)$, we found that $\veps=0.21$ is the best choice for this
material. The result clearly shows that, in order to understand high
- $T_c$ superconductivity, it is necessary to include the
fluctuation contribution as well as the contribution from the
dimensionality of the sample. The method gives a theoretical tool to
estimate the effective dimensionality of the samples.
\end{abstract}
\newpage

\section{Introduction}

 The Ginzburg-Landau (GL) theory of  superconductivity \ci{kl1} had
been  proposed long
 before the famous BCS microscopic theory of superconductivity was discovered.
A few years after the appearance of the BCS theory, Gorkov derived
the GL theory from the  BCS theory \ci{Gorkov}. Since then, the GL
theory has remained as a main theoretical model in understanding
superconductivity. It is highly relevant for the description of both
type -I \ci{paramos} and type II superconductors, even though the
original BCS theory is inadequate to treat both materials. The
success of the GL theory in the study of modern problems of
superconductivity lies on its universal effective  character  in
which
  the details of the microscopic model are
unimportant.

Even at the level of  meanfield approximation (MFA), the GL theory
yields significant information such as the penetration depth ($\cl$)
and the coherence length ($\xi$) of the superconducting  samples.
Many unconventional properties of superconductivity connected with
the break down of the simple MFA has been studied both analytically
 \ci{cam1}
and numerically using the GL theory\ci{cam3}.
 Particularly, the fluctuations of the gauge field
 were studied recently by Camarda et. al. \ci{camarda}
and Abreu et. al. \ci{abreu} in the Gaussian approximation of
the  field theory. The effective mass parameters of the
Gaussian effective potential (GEP),
$\Omega$ and $\Delta$ , were interpreted as inverses of the coherent length
$\xi=1/\Omega$ and of the  penetration depth $\cl=1/\Delta$, respectively.

In our previous paper \ci{ourzphys} we have estimated corrections to
the Gaussian effective potential for the $U(1)$ scalar
electrodynamics, which represents the standard static GL model of
superconductivity. Although it has been shown that the correction is
significant in $D=3$ dimensions, it was not large enough to explain
the experimental findings. At the same time, we have investigated
the role of quasi two dimensionality in the high $T_c$
superconductivity, by calculating the Gaussian effective potential
for $D=2+2\veps$. It was found that the dimensional contribution at
the Gaussian approximation level gives the correction in the right
direction, but is not large enough to explain the experimental
data\ci{ourzphys}. However, it is known that fluctuation
contributions are much larger in lower dimensions. Therefore, it is
necessary to investigate whether the post Gaussian correction terms
in $D=2+2\veps$ dimensions provide significant contribution to the
mean field result, in order to understand the layered structure of
the high $T_c$ superconductivity. In the present paper, we study the
role of the post Gaussian contributions in $D=2+2\veps$ dimensions
by using the method developed in \ci{ourzphys}.

The paper is organized as follows: in Section II the GL  action is
introduced and basic equations are derived; in Section III, the
theoretical results for  $D=2+2\veps$ will be compared to existing
high $T_c$ experimental data, so that the role of fractal dimensions
can be discussed.
  In the Appendix we calculated
two and three loop integrals in $D=2+2\veps$ dimensions.

\section{Basic equations for the effective masses}

The  Hamiltonian of the  model and explicit expressions for the
effective potential in Euclidean D-dimensional space were given in
\ci{camarda,abreu,ourzphys}. Here we bring the main points for
convenience. The effective potential, i.e., the free energy density,
$\veff={\cal{F}}/{\cal{V}}$ is defined as \be \veff=-\ds\ln  Z \ee
where the partition function is 
\be Z=
\intdf{\cal{D}}A_T\exp\{-\ds\int d^Dx H+\ds\int d^Dx j\phi
+(\vec{j}_A \vecA)\} \tochka \lab{Z} \ee
 The Hamiltonian density is
given by 
\be H=\half( \vec{\nabla}\times\vecA)^2+\half(
\vec{\nabla}\phi)^2
+\half m^2\phi^2+\lambda\phi^4   +\half e^2\phi^2 A^2
+\dsfrac{1}{2\eta} ( \vec{\nabla}\vecA)^2 \lab{h}
 \ee
 where we have
introduced  a gauge fixing term with the limit $\eta\rightarrow 0$
being taken after the calculations are carried out. Note that, we
are using natural units  employing
 $\xi_0$ (coherence length at zero temperature)
and $T_c$ (critical temperature) as the length and the energy
scales, respectively, introduced by \ci{kleinertbook}: 
\be 
\ba
m\rightarrow{m}\xi_{0}^{-1},\quad \quad  x\rightarrow{x}\xi_{0},
\nwl e^2\rightarrow{e^2}\xi_{0}^{-1}T_{c}^{-1}  ,\quad
 \lambda\rightarrow{\lambda}\xi_{0}^{-1}T_{c}^{-1}
 \tochka
\lab{scale}
\ea
\ee

Using the method introduced in refs. \ci{ourzphys,oursingap,yee} one
finds following effective potential: 
\be 
\veff=V_G+\Delta V_G
 \ee
where $V_G$ is the Gaussian part: 
\bea
 V_G
&=&I_1(\Omega)+\half  I_1(\Delta)+\half m^2 \phiz^2+\lambda\phiz^4
+
\half
\inolm[m^2-\om+6\lambda\inolm+12\lambda\phiz^2] \nwl \nonumber
&+&\inold[-\delnol+e^2\inolm+e^2\phiz^2],\lab{vg} \eea
 and $\Delta V_G$ is the correction part:
\bea
\nonumber
\Delta V_G&=& [ - {\displaystyle \frac {1}{2}}  e^{4}
\mathrm{I_2}(\Delta ) - 18 \mathrm{I_2}(\Omega ) \lambda ^{2}
] {\phi _{0}}^{4}
 + \{
 - 3 \lambda  \mathrm{I_2}(\Omega )
[ - \Omega ^{2} + m^{2}
 + 2 \mathrm{I_0}(\Delta ) e^{2} + 12 \lambda  \mathrm{I_0}
(\Omega )]
\nwl
\nonumber
 &-& e^{2} \mathrm{I_2}(\Delta ) [ - \Delta ^{2} + e^{
2} \mathrm{I_0}(\Omega )]
\mbox{} - 8 \lambda ^{2} \mathrm{I_3}(\Omega,\Omega )
-
 {\displaystyle \frac {2}{3}}  e^{4} \mathrm{I_3}(\Delta ,
\Omega )\}{\phi _{0}}^{2}
- {\displaystyle \frac {1}{8}}  \mathrm{I_2}(\Omega )
[ - \Omega ^{2} + m^{2} + 2 \mathrm{I_0}(\Delta ) e^{2}
\nwl
&+& 12
\lambda  \mathrm{I_0}(\Omega )]^{2}
 - {\displaystyle \frac {1}{
2}}  \mathrm{I_2}(\Delta ) [- \Delta ^{2} + e^{2} \mathrm{
I_0}(\Omega )]^{2}
 -
  {\displaystyle \frac {1}{12}}
 e^{4} \mathrm{I_4}(\Delta ,  \Omega ) - {\displaystyle
\frac {1}{2}}  \lambda ^{2} \mathrm{I_4}(\Omega,\Omega )\tochka
\lab{vcor}
\eea
In the above following integrals are introduced:
 \be
 \ba
I_0(M)=\dsint\dsfrac{d^Dp}{(2\pi)^D}\dsfrac{1}{(M^2+p^2)}, \quad
 I_1(M)=\dsfrac{1}{2}\dsint\dsfrac{d^Dp}{(2\pi)^D}\ln(M^2+p^2),
\nwl 
I_2(M)=\dsfrac{2}{(2\pi)^{D}}\dsint \dsfrac{d^D
k}{(k^2+M^2)^2}, \nwl
 I_3(M_1,M_2)=\dsfrac{1}{(2\pi)^{2D}}\dsint \dsfrac{d^D k  d^D p}
{
(k^2+M_{1}^{2})
   (p^2+M_{1}^{2}) ((k+p)^2+M_{2}^{2})         }\vergul
\\
\nwl
I_4(M_1,M_2)=\dsfrac{1}{(2\pi)^{3D}}
\dsint \dsfrac{d^D k  d^D p  d^D q}
{
(k^2+M_{1}^{2})
   (p^2+M_{1}^{2})(q^2+M_{2}^{2})
 }
\dsfrac{1}{
((k+p+q)^2+M_{2}^{2})
}
\tochka
\lab{integs}
\ea
\ee
For $D=3-2\varepsilon$,  these integrals were calculated in
 dimensional regularization in ref.
\ci{integs} and  for $D=2+2\veps$ in the Appendix of the present paper.

The parameters $\Omega$ and $\Delta $ are determined by
the  principle of minimal sensitivity
(PMS):
\small
\bea
\nonumber
\dsfrac{\dd \veff}{\dd\Omega}& =&
0.0362 {\displaystyle \frac {\lambda ^{2}}{
\varepsilon ^{2}}}  + \displaystyle \frac {\lambda}{\varepsilon }
[0.075(\omb ^{2}-m^{2})
 -0.108 \lambda(\lnom+1)- 0.911 {\phizb}
^{2} \lambda
 \nwl\nonumber
&+ &0.145 \lambda ^{2} \ln^{2}\dsfrac{\mu^2}{\omb^2}
+ [(0.290 + 1.823 {\phizb}^{2}) \lambda ^{2}
 +0.151 \lambda(m^{2}-\omb ^{2} )]\lnom
+ 0.064\lambda ^{2}(1+3{\phizb}^{2})^2
\nwl\nonumber
&+&(m^{2}-\omb ^{2})[
0.039 (m^{2}-\omb ^{2})
 +0.954 \lambda{\phizb}^{2}
0.151\lambda]  - 0.108 \lambda ^{2} \ln^{3}\dsfrac{\mu^2}{\omb^2}
 \nwl
& +&
[ 0.113 (\omb ^{2}-m^{2})\lambda- \lambda ^{2}(  0.326
+ 1.367 {\phizb}^{2}) ]
\ln^{2}\dsfrac{\mu^2}{\omb^2}
 +
[
(\omb ^{2}-m^{2})(0.954 \lambda {\phizb}^{2}
\nwl\nonumber
&+&
0.227 m^{2}) \lambda
  - 0.039 (\omb ^{2}-m^{2}))
 -\lambda ^{2} (0.133+5.729 {\phizb}^{4} + 3.215 {\phizb}^{2})
]{\lnom}
\nwl\nonumber
& - &\lambda ^{2} (0.960 {\phizb}^{2}+5.72 {\phizb}^{4}+0.144)
 + \varepsilon
[( \omb ^{2} - m^{2})(0.954 \lambda  {\phizb}^{2}-
0.062  \lambda- 0.039 (\omb ^{2} - m^{2})
]
+O(\veps^2)
=0\quad;
\lab{gapeq}
\eea
\bea
\nonumber
\dsfrac{\dd \veff}{\dd\Delta}& =&
\displaystyle \frac {(0.334 \omb ^{2}
- 0.319 \lambda)}{\varepsilon }
  + ( 0.639 \lambda - 0.334 \omb ^{2}
 ) {\lnom}
  + ( 0.319 \lambda- 0.334 \omb ^{2}  ) {\lndel}
 + (4.015 \lambda  - 4.205 \omb ^{2}) {\phizb}^{2}
 \nwl\nonumber
 &-& 1.003 \omb ^{2} + 0.334 m^{2}
+
\varepsilon \{(0.167 \omb ^{2}- 0.479 \lambda  )
\ln^{2}\dsfrac{\mu^2}{\omb^2}
\nwl\nonumber
 &+ &[(0.334 \omb ^{2} - 0.639 \lambda ) {\lndel} +
1.003 \omb ^{2}
- 0.334 m^{2} - 4.015 \lambda  {\phizb}^{
2}] {\lnom}
\nwl
 &+& [(4.205 \omb ^{2} - 4.015 \lambda ) {\phizb}^{2
} + 0.334 \omb ^{2} - 0.334 m^{2}] {\lndel}
- 0.262
\lambda  - 4.943 \omb ^{2} \delb ^{2}
 \nwl\nonumber
 &+& 8.410 \omb ^{2} {\phizb}^{2} + 0.275 \omb ^{2}
\}+O(\veps^2)=0  \vergul
\eea
\normalsize
where we denote optimal values of $\Omega $
and $\Delta$ by $\omb$ and $\delb$, respectively, and
$\phizb$ is  a stationary point defined from the equation:
\small
\bea
\nonumber
\dsfrac{\dd \veff}{\dd\phi_0}&= &
 {\displaystyle \frac
 {
 \lambda  \omb ^{2}
 \delb ^{2} ( 0.456 \lambda  - 0.477 \omb ^{2})
}
{\varepsilon }}
 + (0.477 \omb ^{2} - 0.911 \lambda ) {\lnom}
 - ( 5.729 \lambda  - 2 \omb ^{2}) {\phizb}^{
2}
\nwl
 &+& 0.477 (\omb ^{2}-m^{2}) - 0.119 \lambda  +
{\displaystyle \frac { \omb ^{2} m^{2}}{2\lambda }
}
 + \{[0.683 \lambda  - 0.238 \omb ^{2}]
\ln^{2}\dsfrac{\mu^2}{\omb^2}
\nwl
\nonumber
&+&[5.729 \lambda  {\phizb}^{2} + 0.477 (m^{2} -
 \omb ^{2})
  + 0.239 \lambda ] {\lnom} + 0.240 \lambda
 - 0.392 \omb ^{2}\}\varepsilon+O(\veps^2)=0  \tochka
\lab{gappsi}
 \eea 
 \normalsize
 In the equations (2.9) - (2.11)
 we have used $\veps$ expansion of the loop
integrals explicitly and numerical values of $\xi_0$, $T_c$  and
$e$. For the cuprate $Tl_2Ca_2Ba_2Cu_3O_{10} (T\cl-2223)$ these
values are \be \ba \xi_0=1.36nm,\quad T_c=121.5K, \quad
e^2=16\pi\alpha k_B T_c \xi_0/\hc =0.0000264. \lab{values} \ea \ee

\section{Results and discussions}

The solutions of the Eqs.   (2.9) - (2.11)
are related to the experimentally measured
GL parameter $\kappa$ as $\kappa=\cl/\xi=\omb/\delb$.
We make an attempt to reproduce  recent experimental data
on $\kappa (T)$  \ci{exper} for high -$T_c$ cuprate superconductor
$Tl_2Ca_2Ba_2Cu_3O_{10} (T\cl-2223)$.

For  this purpose,  we adopt usual linear  $T$ dependence of parametrization
of $m$ and $\lambda$ as:
\be
\ba
m^2=\mnol(1-\tau)+\tau \mc\vergul
\quad
\lambda=\lambda_0(1-\tau)+\tau \lambda_c\vergul
\quad
\tau=T/T_c\vergul
\lab{param}
\ea
\ee
and calculate $\kappa$ by solving  nonlinear equations
(2.9) - (2.11). Due to the parametrization \re{param},
 the model has in general
six input parameters: $\mnol$, $\lambda_0$,  $\mc$, $\lambda_c$,
$\xi_0$ (coherent length) and  $T_c$  (the critical temperature).
The experimental values for the cuprate $T\cl-2223$ are
$\xi_0=1.36nm$ and $T_c=121.5K$. To determine other four parameters
we used the following strategy. For  each given $\veps$, the
parameters $\mnol$ and $\lambda_0$ are fitted to the experimental
values of $\xi$ and $\cl$ at zero temperature: $\xi_0=1.36nm$,
$\cl_0=163nm$. In dimensionless units, \re {scale}, we have
$\omb_0=\omb(\tau=0)=1$ and $\delb_0=\delb(\tau=0)=\xi_0/l_0=0.0083$
which are used to calculate $\mnol$ and $\lambda_0$ from the coupled
 equations (2.9) - (2.11). This procedure
gives the $\veps$ dependence of $m_{0}^{2}$ which is presented in Fig. 1
(solid line).
As in the case of the Gaussian approximation\ci{ourzphys}, $m^2$
remains positive only for very small values of $\veps$, although
nonlinearity produces several $m^2=0$ solutions in this case. We
believe that this smallness again indicates the reliability of the
present post Gaussian approximation method.

The parameters $\mc$
and $\lambda_c$ are fixed in the similar way for each given $\veps$.
 Actually the quantum fluctuations
shift $\mc$ from its zero value given by MFA. On the other hand,
the exact experimental values of $\mc$ and $\lambda_c$ are
unknown, since the GL parameter at  $T=T_c$  is poorly determined.
For this reason, we used the experimental values of $\xi_c$ and
$\ell_c$ at very close points to the critical temperature,
$\tau_c=0.98$ which corresponds to
$\omb_c=\omb(\tau_c)=1/\xi_c=0.128$ and
$\delb_c=\delb(\tau_c)=1/\cl_c=0.0043$  ($\kappa_c=29.6$). Then
solving the equations (2.9) - (2.11)
 numerically
with respect to $m_c$ and $
\lambda_c$, we fix   these   parameters.

After having fixed the input parameters, the temperature dependence
of $\omb(\tau)$, $\delb(\tau)$ as well as the GL parameter $\kappa=\omb(\tau)/\delb(\tau)$
are established by solving  the gap equations   numerically for each
$\veps$. Clearly, the solutions of nonlinear gap equations are not unique. In numerical calculations
we separated  the physical  solutions by observing the sign of $\phizb^2$, which should be positive
 and
that the effective potential at the stationary point
 $\veff(\phizb)$ should has
 a real minimum at this point.
 For $\veps\geq 0.1 $, there is a possibility to adjust $\veps$
to the  recent experimental data on $\kappa (T)$  \ci{exper} for
high -$T_c$ cuprate superconductor $Tl_2Ca_2Ba_2Cu_3O_{10}
(T\cl-2223)$. Our calculations show that, the best choice of
$\veps$ is found to be $\veps=0.21$. The appropriate $\kappa
(\tau)$ is presented in Fig. 2 (solid line).
The dashed line in this figure shows $\kappa (\tau)$ for $D=3$. This
fitting process allows us to get an estimation on the effective
dimensionality of the high - $T_c$ superconducting materials.

\section{Summary}

In the present paper, we have carried out two and three loop
calculations on the Ginzburg-Landau effective potential beyond the
Gaussian approximation for $D=2+2\veps$ fractal dimensions. The
result clearly shows that the higher order corrections are
substantially large to explain the existing experimental data.

This result strongly suggests that in order to explain the
experimental data on high - $T_c$ superconductivity it is necessary
to include the fluctuation contribution as well as the contribution
from the quasi two dimensionality. We have found that the GL
parameter is rather sensitive to $\veps$ when the loop corrections
to the simple Gaussian approximation are taken into account.
 The optimal value of $\veps$ for the cuprate $(T\cl-2223)$
is $\veps=0.21$. It would be interesting to estimate optimal
$\veps$ in fractal dimensions for other cuprates also.

It is to be noted that we have  calculated two and three loop
integrals in $D=2+2\veps$ dimensions using the method of dimensional
regularization.

\section*{Acknowledgments}
We appreciate Prof. J.H. Kim for the useful discussions. A.M.R. is
indebted to the Yonsei  University  for hospitality
 during his stay, where   the main part of
this work was  performed. This research was
 supported in part by BK21 project and
 by Korea Research Foundation under project
numbers KRF-2003-005-C00010 and  KRF-2003-005-C00011.

\def\theequation{A.\arabic{equation}}
\setcounter{equation}{0}
\bc
{\Large\bf Appendix }
\ec
\indent
\section*{Explicit expression for the loop integrals in
$D=2+2\veps$ dimension.}
\setcounter{section}{1}

Here, we consider  the loop  integrals defined in Eqs.
 \re{integs} in  $D=2+2\veps$ dimensions.
In dimensional regularization the integrals $I_0(m)$,  $I_1(m)$ and $I_2(m)$
can be easily calculated in momentum space:
\bea
\nonumber
I_0(m)&=&\dsint\dsfrac{d^Dp}{(2\pi)^D}\dsfrac{1}{(m^2+p^2)}=(\dsfrac
{e^\gamma \mu^2}{4\pi})^{-\veps}
\dsfrac{2\pi^{D/2}        }{\Gamma(D/2) (2\pi)^D       }
\intbes \dsfrac{k^{D-1}dk      }{(k^2+m^2)      }
\nonumber
\nwl
& =&(\dsfrac
{e^\gamma  x}{4\pi})^{-\veps}\dsfrac{\Gamma(-\veps) }{(4\pi)^{1+\veps} }
=
-\dsfrac{1   }{4\pi   } \{\dsfrac{1}{\veps}-\ln(x)+\veps[\dsfrac{\pi^2}{12}+ \dsfrac{\ln^2(x)}
{2}    ]  +O(\veps^2)                      \}
\nonumber
\nwl
I_1(m)&=&\dsfrac{1}{2}\dsint\dsfrac{d^Dp}{(2\pi)^D}\ln(k^2+m^2)
=
-\dsfrac{m^2   }{8\pi   } \{\dsfrac{1}{\veps}-1-\ln(x)+\
\veps[\ln(x)+\dsfrac{\pi^2}{12}+1
\nonumber
\nwl
&+& \dsfrac{\ln^2(x)}
{2}    ]   +O(\veps^2)                    \}
\nonumber
\nwl
I_2(m)&=&2\dsint\dsfrac{d^Dp}{(2\pi)^D (k^2+m^2)^2        }=
\dsfrac{1   }{2m^2\pi   } \{  1-\veps\ln(x)
+O(\veps^2)                    \}
\vergul
\eea
with $x=\mu^2/m^2$.

Two and three loop integrals ($I_3$ and $I_4$) require a little more
effort. It is more convenient to evaluate them in coordinate space
rather than in momentum space, since

\bea
\nonumber
I_3(M_1,M_2)&=&\dsfrac{1}{(2\pi)^{2D}}\dsint \dsfrac{d^D k  d^D p}{(k^2+M_{1}^{2})
   (p^2+M_{1}^{2}) ((k+p)^2+M_{2}^{2})         }
=(\dsfrac
{e^\gamma \mu^2}{4\pi})^{\veps}\dsint d^D r G_{1}^{2}(r)G_2 (r)
\nonumber
\nwl
I_4(M_1,M_2)&=&\dsfrac{1}{(2\pi)^{3D}}
\dsint \dsfrac{d^D k  d^D p  d^D q}
{
(k^2+M_{1}^{2})
   (p^2+M_{1}^{2})(q^2+M_{2}^{2})
 }
\dsfrac{1}{
((k+p+q)^2+M_{2}^{2})
}
\nwl
\nonumber
&=&(\dsfrac
{e^\gamma \mu^2}{4\pi})^{\veps}\dsint d^D r G_{1}^{2}(r)G_{2}^{2} (r)
\vergul
\lab{integs34}
\eea
where $G_n(r)$ is the Fourier  transform of the propagator $1/(k^2+M_{n}^{2})$ $ (n=1,2)$:
\be
\ba
G(r)=\dsint\dsfrac{d^Dk e^{ikr}   }{(2\pi)^D (k^2+m^2)}=\dsfrac{(2\pi)^{-D/2} m^{D-2}           }
{(mr)^{D/2-1}        } K_{D/2-1}(mr)
\ea
\ee
 and  $K_\nu (z) $ is the modified Bessel function. In dimensional regularization, for
$D=2+2\veps$, $G(r)$ is simplified as \be G(r)=\ds{\ds
(\dsfrac{e^\gamma x}{2})^{-\veps}}\dsfrac{(mr)^{-\veps}
}{2\pi}K_\veps(mr) \tochka \lab{G} \ee Now, substituting \re{G}
into \re{integs34} one notices that unlike in
 the case of $D=3-2\veps$ , in $D=2+2\veps$ dimensions there is no singularity at
small $r$ and hence the   integration  can be performed directly from $r=0$
 to $r=\infty$  without splitting radial integration into two regions with small $r$
and large $r$.

The case with equal  masses, $M_1=M_2\equiv m$,  can be done
analytically:
\be
\ba
I_N(m)=\dsfrac{2^{-N\veps}(e^\gamma x/4)^{\veps(1-N)}           }
             {(2\pi)^{N-1}m^2\Gamma(1+\veps)              } \tilde{I}_N(\veps)
,\quad\quad
\tilde{I}_N(\veps)=\intbes
t^{1+2\veps}[t^{-\veps}K_\veps (t)    ]^N dt \nwl \nwl
 \ea \ee
for $N=3,4$, where the  integrals  $\tilde{I}_3(\veps)$ and
$\tilde{I}_4(\veps)$ are expressed in term of the hypergeometric
functions: 
\bea \nonumber
\tilde{I}_3(\veps)&=&\dsfrac{\Gamma(\veps)\Gamma(1-\veps)}{2^{3+\veps}
} \{\dsfrac{4^\veps\sqrt{\pi} \Gamma(1-2\veps) }{\Gamma(3/2-\veps) }
 { _2{F}_1 }[1,1-2\veps;
 \frac{3}{2}-\veps;\frac{1}{4}      ]
\nonumber
\nwl
\nonumber
\nwl
&-& 2\Gamma(1-\veps) \; { _2F_1} [1,1-\veps;
 \frac{3}{2};\frac{1}{4}      ]                  \},  \quad (\veps\le 0.5);
\nonumber
\nwl
\nwl
\tilde{I}_4(\veps)&=&\dsfrac{\Gamma(\veps)\Gamma(1-\veps)}{8   }
\{\dsfrac{ \veps\Gamma^2(-\veps)   }{  4^\veps }
{ _3{F}_2 }[1,1-\veps,
 \frac{1}{2}+\veps;\frac{3}{2},1+2\veps;1  ]
\nonumber
\nwl
\nonumber
\nwl
&+&
\dsfrac{ 2         \Gamma^2(-\veps) \veps  }{4^{\veps}(  2\veps-1) } \;\; { _3{F}_2 }[
 \frac{1}{2},1,
 1-2\veps; \frac{3}{2}-\veps,1+\veps;1  ]
\nonumber
\nwl
\nonumber
\nwl
\nonumber
&-&
\dsfrac{4^\veps \sqrt{\pi} \Gamma(1-3\veps)\Gamma(1+\veps)\Gamma(-2\veps)   }
{\Gamma(\frac{3}{2}-2\veps) }
{ _2{F}_1 }[
 1-3\veps,\frac{1}{2}-\veps;
\frac{3}{2}-2\veps;1  ]
  \}
 ,\quad (\veps\le 1/3) \tochka
\eea

 The method of ref. \ci{davyd} gives the following $\veps$
 expansion:
\be
\ba
I_3(m)=\dsfrac{1}{4\pi^2 m^2}[0.5917+\veps(0.6629-1.1835 \ln x)+O(\veps^2)]\vergul
\nwl
I_4(m)=\dsfrac{1}{8\pi^3 m^2}[1.188-\veps(2.759+3.5656\ln x)+O(\veps^2)]
\ea
\ee
which is used in our practical calculations.

The case with nonequal masses is rather complicated and cannot be
done analitically in general.  However, in the particular case, when
$\alpha\equiv M_2/M_1<1 $ \footnote{in the present paper
$\alpha=1/\kappa$ where $\kappa\approx 80 $ in the large range of
temperature} the problem may be overcome by expansion in power
series in $\alpha$. We shall illustrate this approximation for
$I_3(M_1,M_2)$ below. Using  Eq.s \re{integs34} and \re{G} one
obtains 
\bea
I_3(M_1,M_2)&=&(\dsfrac{e^\gamma \mu^2}{4\pi})^\veps\ds\int\;G_{1}^{2}(r)  G_{2}(r) d^D r
=\dsfrac{1 }{4\pi^2M_{1}^{2} \Gamma(\veps+1)}
[\dsfrac{x_1x_2 \exp(2\gamma)}{2   }
        ]^{-\veps} \tilde{I}_3(\alpha,\veps) \vergul
\lab{i3m12}
\eea
where
\be
\tilde{I}_3(\alpha,\veps)=\intbes t K_{\veps}^{2}(t)\;(\alpha t)^{-\veps}K_\veps(\alpha t)
\tochka
\lab{kint}
\ee
Now using the series  expansion  of $K_\nu(z)$
\bea
\nonumber
K_\nu(z)&=&\dsfrac{\Gamma(\nu) \Gamma(1-\nu)    }{ 2         }\{
z^{-\nu}[\dsfrac{2^\nu}{\Gamma(1-\nu)} + \dsfrac{2^{\nu-2}z^2       }{\Gamma(2-\nu)}+O(z^4)
             ]
\nwl
&-&
z^{\nu}[\dsfrac{2^{-\nu}}{\Gamma(1+\nu)} + \dsfrac{2^{-\nu-2}z^2       }{\Gamma(2+\nu)}+O(z^4)
             ]
\}\vergul
\eea
one may expand the factor $(\alpha t)^{-\veps}K_\veps(\alpha t)$ in power series of $\alpha$
and integrate \re{kint} analytically to obtain:
 \bea
\nonumber \tilde{I}_3(\alpha,\veps)&=&
-\dsfrac{\veps^2\Gamma(\veps)\Gamma^2(-\veps) } {24
(2\veps-1)(2\veps-3) 2^\veps  }
 \{
(2\veps-1)(2\veps-3) (\alpha^2 \veps-\alpha^2-6)
\nonumber \nwl & 
-&3 \alpha^{-2\veps}
[4\veps-6+\alpha^2(2\veps-1) ]
+O(\alpha^4)        \}\tochka \lab{i3til} \eea
 Inserting  Eq.
\re{i3til} into  the Eq. \re{i3m12} one obtains the following
$\epsilon$ expansion : 
\bea \nonumber
I_3(M_1,M_2)&=&\dsfrac{1}{864\pi^2 M_{1}^{2} } \{
108(1-\lna)-3\alpha^2(6\lna -5 ) 
+
\veps[\alpha^2(-18\lnaa+(36\lnx+18)\lna
\nonumber \nwl
&-&30\lnx
+
4) - 216\lnx-108\lnaa+216 
+216\lnx\lna
+O(\veps^2)                           ] \}\tochka \eea 
Similarly,
one may  calculate $I_4(M_1,M_2)$  to obtain it's    $\epsilon$
expansion: 
\bea 
\nonumber 
I_4(M_1,M_2)&=&\dsfrac{1}{1728\pi^3
M_{1}^{2} } \{4\alpha^2(2+9\lnaa-6\lna   ) 
-108\lnaa+190.9588\lna-280.5109
 \nonumber \nwl
&+&\veps\;[\alpha^2(72\ln^3\alpha-(60+108\lnx)\lnaa
+ (72\lnx-28)\lna-24\lnx+23.4519)
 \nonumber \nwl
&-&360\ln^3\alpha+(547.8351+324\lnx)\lnaa
-(572.8764\lnx+337.6413)\lna 
\nonumber \nwl
&+&841.5330\lnx
 -806.1519
+O(\veps^2) ] \}
 \eea
 where, for simplicity, we used explicit values
of constants such as $\gamma$, $\zeta(3)$, $\ln(2)$, etc.

\bb{99} \bi{kl1} V. L. Ginzburg and L. D. Landau, Zh. Eksp. Teor.
Fiz. {\bf 20}, 1064, (1950) . \bi{Gorkov}L. P. Gorkov, Sov. Phys.
JETP {\bf 7}, 505 (1958); L. P. Gorkov, Sov. Phys. JETP {\bf 9},
1364 (1959). \bi{paramos}J. Paramos, O. Bertolami, T. A. Girard
and P. Valko,
 Phys. Rev. B{\bf 67}, 134511, (2003).
\bi{cam1} Z. Tesanovich, Phys. Rev. B{\bf 59}, 6449, (1999) (and
references there in). \bi{cam3} A. K. Nguyen and A. Sudbo, Phys.
Rev. B{\bf 60} 15307 (1999). \bi{camarda}  M. Camarda, G. G. N.
Angilella, R. Pucci
 and F. Siringo ,  Eur. Phys. J.  B{\bf 33} , 273, (2003).
\bi{abreu}
 L. M. Abreu, A. P. C. Malbouisson and I. Roditi,
cond-mat/0305366 (15-May 2003)
 \bi{ourzphys} C. K. Kim, A. Rakhimov and  J. H. Yee, Eur. Phys. J.
 \textbf{B39},301(2004)
\bi{kleinertbook}H. Kleinert, {\it Gauge Fields in Condensed Matter}, Vol. 1:
Superflow and Vortex Lines,  World Scientific , Singapore,1989.
\bibitem{oursingap}  A. Rakhimov  and J. H. Yee,  Int. J. Mod. Phys. {\bf A19}, 1589 ( 2004).
\bibitem{yee} G. H. Lee and J. H. Yee,  Phys. Rev. D{\bf 56}, 6573,  (1997).
\bi{integs} E. Braaten and A. Nieto, Phys. Rev. D{\bf 51}, 6990,
(1995); A. K. Rajantie Nucl. Phys. B{\bf  480}, 729, (1996). \nwl
M. Yu. Kalmykov and O. Veretin , Phys. Lett. B{\bf 483}, 315,
(2000) \bi{exper} G. Brandstatter, F. M. Sauerzopf, H. W. Weber,
F. Ladenberger and E. Schwarzmann, Physica C, {\bf 235}, 1845,
(1994); \nwl G. Brandstatter, F. M. Sauerzopf and  H. W. Weber,
Phys. Rev. B{\bf 55}, 11693, (1997). \bi{davyd} A. I. Davydychev
and M. Yu. Kalmykov Nucl. Phys. B{\bf  605}, 266, (2001);
  \eb

\newpage
\begin{figure}
 \epsfxsize=15.cm
\begin{center}
\epsffile{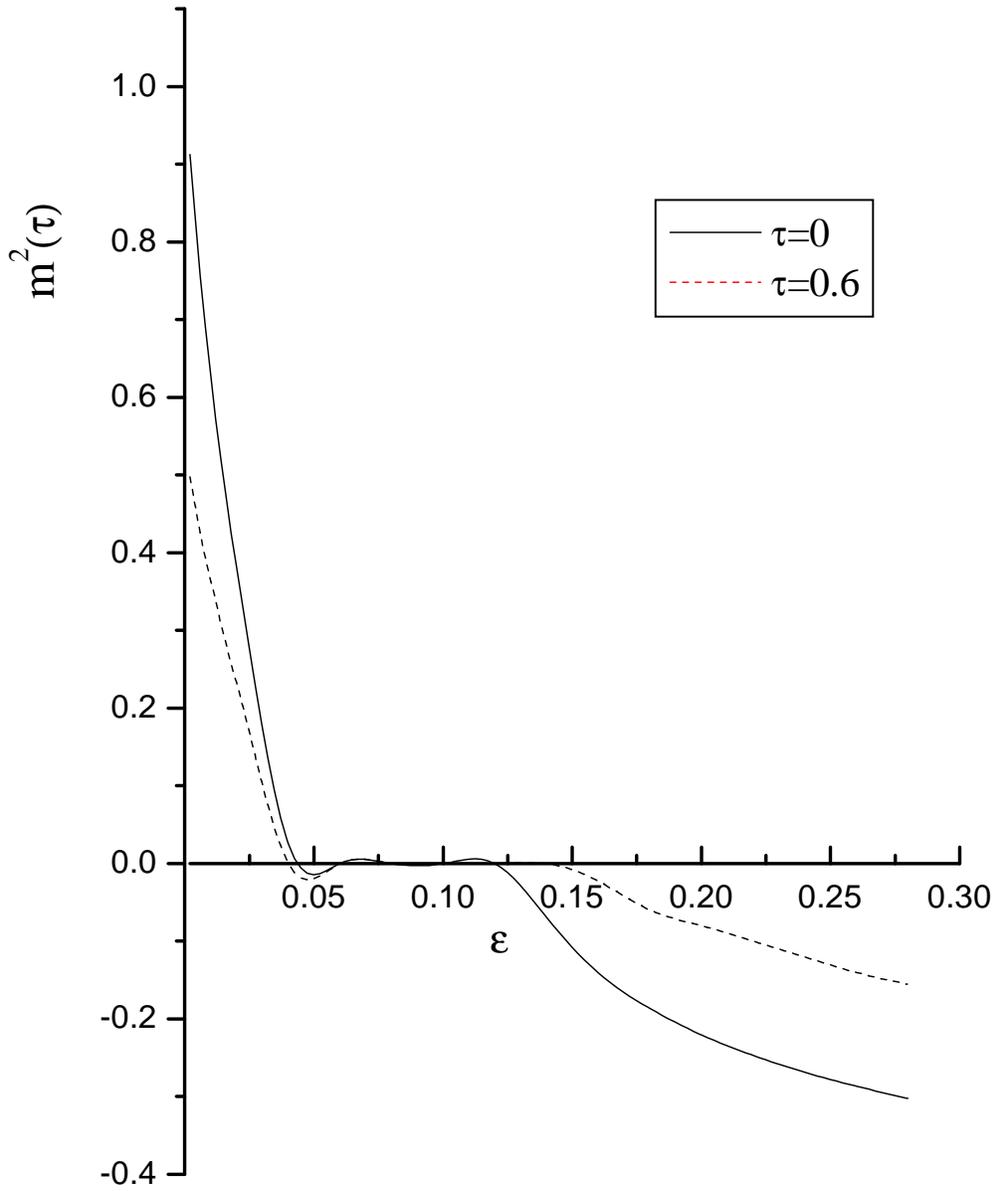}
 \end{center}
\caption{
The parameter $m^2$ of the GL model v.s. $\veps$ in fractal dimension
$D=2+2\veps$. The solid and dashed lines are for the temperatures
$T=0$ and $T=0.6T_c$ respectively.
}
\end{figure}
\newpage

\begin{figure}
 \epsfxsize=15.cm
\begin{center}
\epsffile{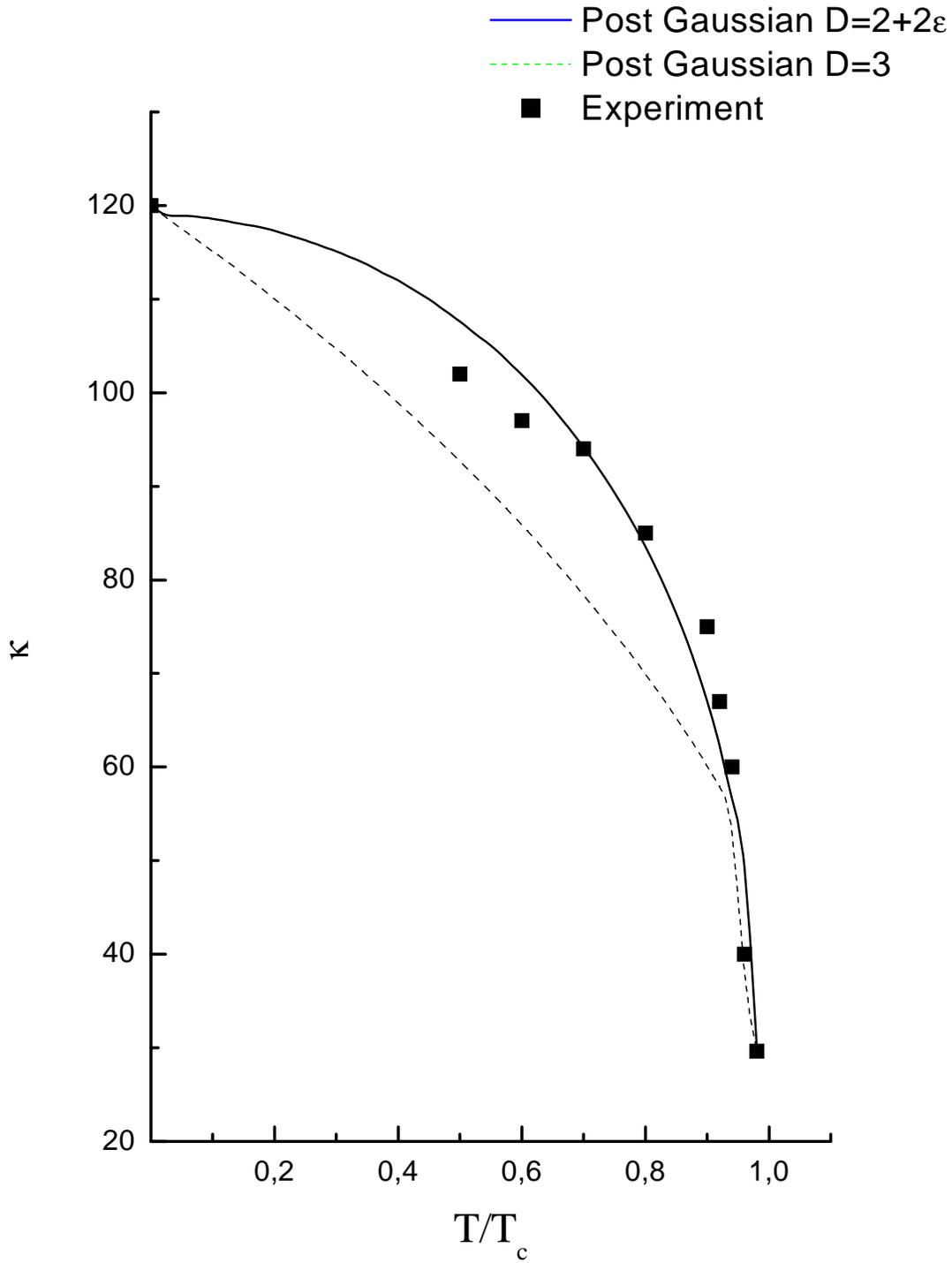}
 \end{center}
\caption{ The GL parameter, $\kappa  $, in $D=2+2\veps$ (solid
line)  and in $D=3$ (dashed line) cases calculated in the
PostGaussian approximation. }
\end{figure}

\edc